\documentclass[twocolumn,showpacs,showkeys,preprintnumbers,amsmath,amssymb]{revtex4}

\usepackage{graphicx}
\usepackage{dcolumn}
\usepackage{bm}

\usepackage[normalem]{ulem} 
\usepackage{color} 

\newcommand{\ket}[1]{\ensuremath{\left| #1 \right\rangle}}
\newcommand{\braket}[2]{\ensuremath{\left\langle #1 | #2\right\rangle}}
\newcommand{\mean}[1]{\ensuremath{\left\langle #1 \right\rangle}}
\newcommand{\var}[1]{\ensuremath{\left\langle \left\langle #1 \right\rangle \right\rangle}}

\begin{document}

\title{Relaxation of Josephson qubits due to strong coupling to two-level systems}

\author{Clemens M\"uller}
\author{Alexander Shnirman}
\affiliation{Institut f\"ur Theorie der Kondensierten Materie and DFG-Center for Functional Nanostructures (CFN), 
Universit\"at Karlsruhe, D-76128 Karlsruhe, Germany}

\author{Yuriy Makhlin}
\affiliation{Landau Institute for Theoretical Physics, Kosygin Street. 2, 119334, Moscow, Russia}

\date{\today}

\begin{abstract}
	We investigate the energy relaxation ($T_1$) process of a qubit coupled
	to a bath of dissipative two-level fluctuators (TLFs). We consider the 
	fluctuators strongly coupled to the qubit both in the limit of spectrally sparse single 
	TLFs as well as in the limit of spectrally dense TLFs. We conclude that the 
	avoided level crossings, usually attributed to very strongly coupled single 
	TLFs, could also be caused by many weakly coupled spectrally dense fluctuators.
\end{abstract}


\maketitle

\section{\label{sec:Introduction} Introduction}
	
With the current progress in fabrication, manipulation, and measurement of superconducting qubits
it became crucial to understand the microscopic nature of the environment, responsible for decoherence. 
Recent experiments showed considerable advances even in the presence of the low-frequency, $1/f$, noise without eliminating its sources. 
The operation at the `optimal point' in the {\it quantronium}~\cite{Ithier05} or the drastic reduction 
of unharmonicity in the {\it transmon}~\cite{Koch07} reduced dephasing (increased $T_2$) significantly.
Especially in the transmon the relation $T_2 \approx 2 T_1$ is now
routinely reported, which implies that the decoherence is limited by the energy relaxation ($T_1$) process, i.e., the pure dephasing is suppressed.

The microscopic nature of the dissipative environment, responsible for the energy relaxation, is still unknown.
While the effects of the electromagnetic environment, including the Purcell effect, can be reliably 
estimated~\cite{houck:080502}, the intrinsic sources of relaxation remain unidentified. 
There exist numerous indications that the charge and the critical-current noise are induced by collections of two-level fluctuators (TLFs) residing, 
possibly, in the tunnel junction or at the surface 
of the superconductors~\cite{Astafiev04,harlingen:064517}. 
Several microscopic models of TLFs have been proposed~\cite{Faoro123,faoro:047001,koch:267003,faoro:132505}.
It was suggested~\cite{Astafiev04,Shnirman05} that in charge qubits the relaxation is due to charge 
fluctuators which are simultaneously responsible for the $1/f$ charge noise~\cite{Dutta81}.
On the other hand, the flux noise might also be due to a large number of 
paramagnetic impurities~\cite{faoro:227005,Sendelbach08}. 
In addition, strong signatures of resident two-level systems were found, especially in Josephson phase qubits~\cite{Simmonds04,Cooper04}, in which the Josephson junctions have relatively large areas. Recently the coherent dynamics of a qubit 
coupled strongly to a single TLF was explored with the idea to use the TLF 
as a naturally formed quantum memory~\cite{Neeley08,Zagoskin06}.
	
Strong coupling to fluctuators as a source of decoherence of a qubit was the focus of 
research in the past, cf., e.g., Refs.~\onlinecite{Paladino02,deSousa05,Grishin05}. These works concentrated,
however, on single-level fluctuators, i.e., an electron jumping back and forth between a continuum
and a localized level. 
Such a system maps onto an overdamped dissipative two-level 
system~\cite{Weiss}. In contrast, here we study the effect of strong coupling to 
underdamped (coherent) two-level fluctuators.

In this paper we analyze the properties of a qubit coupled to TLFs and, in particular, the relaxation process.
On one hand, we study the dissipative dynamics of a qubit strongly coupled to a single fluctuator; these results could be useful 
for the analysis of the experimental data, e.g., for phase qubits. 
On the other hand, we investigate a qubit coupled to a collection of TLFs, and our findings allow us to speculate about some general features of the microscopic picture behind the phenomenon of qubit relaxation, which could account for the experimental observations. In the next section we discuss in more detail the analysis in this paper and possible features of the fluctuator bath.

\section{Motivation and discussion}

Below we investigate the dynamics of a qubit coupled to one or many TLFs. Our findings, together with certain observations concerning the experimental data, allow us to speculate about the properties of the collections of TLFs in real samples. In other words, we suggest a possible structure of the fluctuator bath, which is consistent with these observations.

In this paper, we first consider a single TLF coupled resonantly to the qubit, i.e., with its energy splitting close to that of the qubit. These results can be useful for the analysis of experiments, in which the qubit is coupled resonantly to a single TLF (while the other fluctuators are far away from resonance and do not contribute to the qubit's dynamics) or strongly coupled to one TLF and weakly to the rest, which form a `background'.
In this regime the TLF strongly affects the qubit's dynamics, and we observe two effects: (i) coherent oscillations with the excitation energy going back and forth between the qubit and the TLF; and (ii) the decay to the ground state due to the energy relaxation in either the TLF or the qubit. The oscillations themselves also show decay, dominated by dephasing processes. We describe the oscillation and relaxation processes and determine the relevant time scales. 

Further, we discuss the dynamics of a qubit coupled to a collection of TLFs.
Our motivation is based on the following observations from the analysis of the experimental data: 
(a) strongly coupled TLFs ({\it strongly coupled} refers to a strong qubit-TLF coupling) were observed experimentally in phase qubits with large-area 
junctions~\cite{Simmonds04,Cooper04,Neeley08}. In these qubits the $T_1$ time 
shows rather regular behavior as a function of the energy splitting of the qubit (in the regions between 
the avoided level crossings, which arise in resonance with the strongly coupled TLFs);
(b) in smaller phase and flux qubits the $T_1$ time often shows a seemingly random 
behavior as a function of the energy splitting of the 
qubit~\cite{Astafiev04,Ithier05,NakamuraPrivate}; 
(c) the strong coupling observed in Refs.~\onlinecite{Simmonds04,Cooper04,Neeley08} requires a microscopic explanation. For instance, a large dipole moment of the TLF, $e d$, is needed to account for the data, where
$d$ is of the order of the width of the tunnel barrier and $e$ is the electron charge; 
and (d) experiments~\cite{Sendelbach08} 
suggest a very high density of (spin) fluctuators on the surface of superconductors. 

Based on these observations we speculate about a possible microscopic picture of the fluctuations, which could be consistent with these observations: First, one could expect in the analysis of the dependence of $T_1$ on the level splitting that the contribution of each fluctuator is peaked near its level splitting (when it is resonant with the qubit and can absorb its energy efficiently). Further, one might assume that for a large collection of spectrally dense TLFs (that is with a dense distribution of the level splittings), the corresponding peaks overlap strongly, and the resulting $T_1$-energy curve is smooth (even though for a dense distribution the contributions of the TLFs are not necessarily independent). Indeed, this general picture is consistent with the data:
in charge and flux qubits, with smaller-area junctions, the TLFs are not 
spectrally dense, and resonances with single TLFs can be resolved in the dependence of the relaxation rate on the level splitting. This may look as a seemingly random collection of peaks.
In contrast, in phase qubits, with large-area junctions, there are many TLFs (for instance, the TLFs could be located in the junctions so that their number would scale with the junction area); thus the spectral distribution of their level splittings is dense and almost continuous. This may produce a smooth $T_1$-vs.-energy curve.
Furthermore, one can speculate about the structure of the fluctuator bath. Suggested scenarios of the microscopic nature of the fluctuators find it difficult to explain the existence of the strongly coupled TLFs, which were observed, for instance, in the qubit spectroscopy via the avoided level crossings~\cite{Simmonds04,Cooper04,Neeley08}. 

In other words, in our picture each TLF is only coupled to the qubit, and they are essentially decoupled from each other. 
For each of them the coupling to the qubit is much weaker
than observed in experiments~\cite{Simmonds04,Cooper04,Neeley08}.
For a dense uniform distribution of the TLFs splittings, usual relaxation of the qubit takes place.
However, as we find below, if the level splittings of the TLFs accumulate close to some energy value (which may be a consequence of the microscopic nature of the TLFs), as far as the qubit's dynamics is concerned the situation is equivalent to a single strongly coupled TLF. Thus, in our picture, 
weakly coupled TLFs may conspire to emulate a strongly coupled TLF, visible, e.g., via qubit spectroscopy. 
Note, however, also the results of 
Ref.~\cite{Lupascu2008,HofheinzPrivate,LisenfeldPrivate} pointing towards single strongly coupled TLFs.

To demonstrate this kind of behavior, we further study the regime, where two or more fluctuators are in resonance with the qubit. 
Our main observation in this case is that the fluctuators form a single effective TLF with 
stronger coupling to the qubit. 

For a collection of many TLFs with a low spectral density, we estimate the statistical characteristics (by averaging over the possible spectral distributions) of the random relaxation rate of the qubit and estimate corrections to this statistics due to the resonances that involve multiple TLFs.

Finally, we discuss collections of spectrally dense TLFs. In this case we identify two regimes.
If the TLFs are distributed homogeneously in the spectrum, they form a continuum, to which the qubit 
relaxes, and the dynamics is described by a simple exponential decay. If, however, a sufficiently 
strong local fluctuation of the spectral density of TLFs occurs, the situation resembles again 
that with a single, strongly coupled fluctuator. This may explain the origin of the strongly coupled TLFs observed in the experiment.

\section{\label{sec:System}Model}

We consider the system described by the following Hamiltonian
\begin{equation}
	\hat{H} = -\frac{1}{2}\: \epsilon_q\: \sigma_z - 
		\frac{1}{2}\:\sum_{n}\: \epsilon_{f,n}\: \tau_{z,n} + \frac{1}{2}\: \sigma_x\: \sum_{n} v_{\perp,n}\: \tau_{x,n} + \hat{H}_{\rm Bath}\ .
	\label{eq:Hmain}
\end{equation}
The first term, the Hamiltonian of the qubit in its eigenbasis reads
$\hat{H}_q = -\frac{1}{2}\: \epsilon_q\: \sigma_z$, 
where $\epsilon_q$ is the level splitting between 
the ground and the excited states, and $\sigma_z$ is the Pauli matrix. Similarly, the Hamiltonian of the
$n$-th TLF in its eigenbasis reads $\hat{H}_{f,n} = -\frac{1}{2}\: \epsilon_{f,n}\: \tau_{z,n}$.
We consider only transverse couplings with the strengths $v_{\perp,n}$, described by the third term. 
We note that the qubit-TLF interaction (e.g., the charge-charge coupling) would typically produce also other coupling terms in the qubits's eigenbasis
(longitudinal and mixed terms; cf. the discussion of the purely longitudinal coupling $\propto \sigma_{z} \tau_{z}$ relevant for the dephasing by $1/f$ noise,
e.g., in Refs~\cite{Galperin:2006p72, Schriefl:2006p35}). However, for our purposes (description of relaxation) the transverse coupling is most relevant since it gives rise to spin-flip processes between the qubit and TLFs.
In our model all TLFs interact with the qubit, but not with each other. This assumption is reasonable, 
since the TLFs are microscopic objects distributed over, e.g., the whole area of the 
Josephson junction. Thus, they typically are located far from each other, but interact
with the large qubit. The term $\hat{H}_{\rm Bath}$ describes the coupling of each TLF and of the qubit 
to their respective baths. We model the environment of the qubit and of the TLFs as a set of baths 
characterized by the variables $X_i$ and coupling constants $\beta_i$ (specific examples are provided below). We write down and 
solve the Bloch-Redfield equations~\cite{Bloch57,Redfield57} for the coupled system of qubit and TLFs. 

In the course of solving the Bloch-Redfield equations many rates (elements of the Redfield tensor)
play a role. It is useful to reduce those rates, when possible, to the `fundamental' rates, i.e., those 
characterizing the decoupled TLFs and the qubit.
Each fluctuator is thus characterized by its own relaxation rate $\gamma^{f,n}_1$
and by the pure dephasing rate $\gamma^{f,n}_\varphi$, with the total dephasing rate 
given by $\gamma^{f,n}_2 = (1/2) \gamma^{f,n}_1 + \gamma^{f,n}_\varphi$. We define these 
rates below and also discuss the generalization for the case of a non-Markovian environment. 
Also the qubit is characterized by its intrinsic (not related to fluctuators) relaxation rate $\gamma^{q}_1$ 
and the pure dephasing rate $\gamma^{q}_\varphi$. Again $\gamma^{q}_2 = 
(1/2) \gamma^{q}_1 + \gamma^{q}_\varphi$. In what follows the mentioned rates are treated 
as fundamental. All the other rates, emerging in the coupled system of the qubit and fluctuators,
are denoted by capital letters $\Gamma$.

\section{\label{sec:SingleTLF}Single TLF}

\begin{figure}[ht]
	\includegraphics[width=\columnwidth]{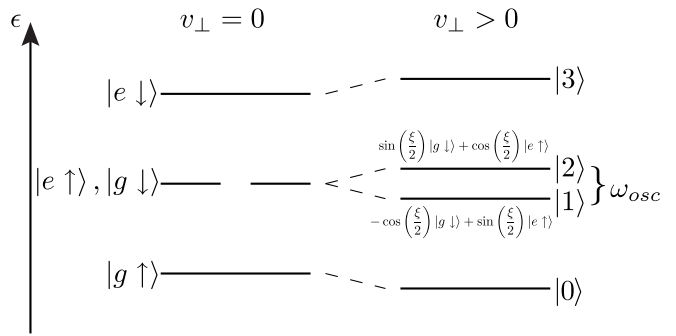}
	\caption{Level structure of the coupled qubit-TLF system in resonance $\delta\omega=0$. 
		For $v_{\perp}=0$ the middle levels form a degenerate doublet. The coupling lifts the
		degeneracy and splits the levels by the oscillation frequency $\omega_{osc}$.}
	\label{fig:Levels}
\end{figure}

We first consider a system of a qubit and a single TLF.
\begin{equation}
	\hat{H} = -\frac{1}{2}\: \epsilon_q\: \sigma_z - 
		\frac{1}{2}\: \epsilon_f\: \tau_z + \frac{1}{2}\: v_{\perp}\: \sigma_x\: \tau_x + \hat{H}_{\rm Bath} \ .
\label{eq:H}
\end{equation}
We restrict ourselves to the regime $\epsilon_q \approx \epsilon_f \gg v_{\perp}$. Then the 
ground state $\ket{0}\approx\ket{g\uparrow}$ and the highest energy level 
$\ket{3}\approx\ket{e\downarrow}$ are only slightly affected by the coupling.
On the other hand the states $\ket{e\uparrow}$ and $\ket{g\downarrow}$
form an almost degenerate doublet. The coupling $v_\perp$ lifts the degeneracy to form the two 
eigenstates $\ket{1}=-\cos\frac{\xi}{2}\: \ket{g\downarrow} + \sin\frac{\xi}{2}\: \ket{e\uparrow}$ and 
$\ket{2}=\sin\frac{\xi}{2}\: \ket{g\downarrow} + \cos\frac{\xi}{2}\: \ket{e\uparrow}$ (cf. Fig.~\ref{fig:Levels}).
Here we introduced the angle 
$\tan{\xi} = v_\perp/\delta\omega$ where $\delta\omega\equiv\epsilon_q-\epsilon_f$ 
is the detuning between the qubit and the TLF. 
The energy splitting between the levels $\ket{1}$ and $\ket{2}$ is given by 
$\omega_{osc} = \sqrt{v_\perp^2+\delta\omega^2}$.

\subsection{\label{sub:1TLFa}Transverse TLF-bath coupling}

	First, we consider the simplest case, in which only the TLF is coupled to a dissipative bath 
	and this coupling is transverse. The coupling operator in Eq.~(\ref{eq:H}) takes the form
	\begin{equation}
		\hat{H}_{\rm Bath} = \frac{1}{2} \beta_{f,\perp}\: \tau_x \cdot X_{f,\perp}\ ,
		\label{eq:Hi1}
	\end{equation}
	where the bath variable $\hat{X}_{f,\perp}$ is characterized by the (non-symmetrized) correlation function 
	$C_{f,\perp}(t) \equiv \mean{ \hat{X}_{f,\perp}(t) \hat{X}_{f,\perp}(0)}$. In thermal equilibrium we have 
	$C_{f,\perp}(-\omega)=e^{-\omega/T} C_{f,\perp}(\omega)$. We assume here that $T \ll \epsilon_f$, i.e., 
	that the temperature is effectively zero, so that we can neglect excitations.
	
\begin{figure}[ht]
		\includegraphics[width=\columnwidth]{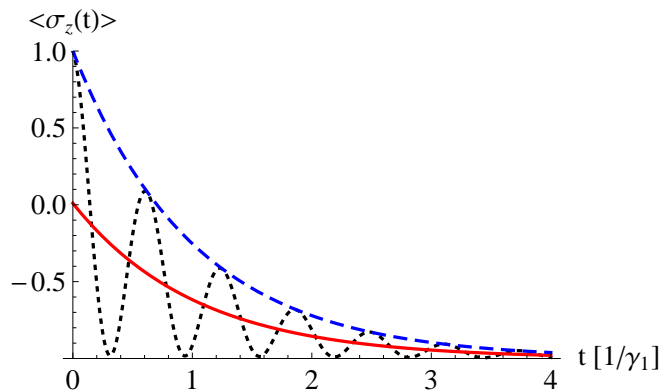}
		\caption{$\mean{\sigma_z}$ as a function of time in units of the inverse TLF relaxation rate $\gamma_{1}^f$ 
			for the case of the qubit exactly in resonance with the TLF (dotted black line). 
			One observes oscillations with frequency $\omega_{osc}$. The solid red curve gives the decay averaged 
			over the oscillations, characterized by $a_{av}$ and $\Gamma_{av}$ 
			and the dashed blue curve shows the envelope described by $a_{env}$ and $\Gamma_{env}$.
			Parameters in this plot are (in units of $\gamma_{1}^f$): $\epsilon_{q}=\epsilon_{f}=100, v_{\perp}=10$}.
		\label{fig:SigmaMean}
	\end{figure}

	We solve the Bloch-Redfield equations~\cite{Bloch57,Redfield57} for the coupled system using the secular approximation. 
	As the initial condition we take the qubit in the excited state and the TLF in its thermal equilibrium state.
	Tracing out the TLF's degrees of freedom we find the dynamics of 
	$\mean{\sigma_z}$ (Fig.~\ref{fig:SigmaMean}). 
	
	For the expectation value $\mean{\sigma_z}$ we find the following expression
	\begin{equation}
		\begin{split}
			\mean{\sigma_z(t)} =& \mean{\sigma_z}_\infty + a_{\downarrow, 1}
			\: e^{-\Gamma_{\downarrow, 1}\: t} + a_{\downarrow, 2} \: e^{-\Gamma_{\downarrow, 2}\: t}\\
			+& a_{osc}\: \cos{\left( \omega_{osc}t \right)} e^{-\Gamma_{osc}\: t}\ ,
		\end{split}
		\label{eq:SigmaA}
	\end{equation}
	where $\mean{\sigma_z}_\infty \approx -1$ is the zero-temperature equilibrium value. 
	We can separate the right-hand side of Eq.~(\ref{eq:SigmaA}) into damped oscillations, with decay
	rate $\Gamma_{osc}$, and a purely decaying part. The amplitude and the decay rate of the oscillating 
	part are given by
		\begin{eqnarray}
			a_{osc} &=& \frac{v_\perp^2}{v_\perp^2+\delta\omega^2}
			\label{eq:aOscA}\ ,\\
			\Gamma_{osc} &=& \frac{1}{2}\: \gamma_1^f\ .
			\label{eq:GOscA}
		\end{eqnarray}
		Here the rate 
	\begin{equation}
		\gamma_1^f = \frac{1}{4}\: \beta_{f,\perp}^2\: C_{f,\perp}(\omega\approx\epsilon_q)\ .
		\label{eq:gammaT}
	\end{equation}	
	is the relaxation rate of the fluctuator. We observe that the decay rate of the oscillations,
	$\Gamma_{osc}$, is independent of the coupling strength $v_\perp$ and of the 
	detuning $\delta\omega$. 		
	Note that the physics considered here is only relevant near the resonance $\epsilon_f \approx \epsilon_q$,
	and we assume that the spectrum $C_{f,\perp}(\omega)$ is sufficiently smooth in this region, so that $C_{f,\perp}(\epsilon_f)\approx 
	C_{f,\perp}(\epsilon_q)$.
	
\begin{figure}
		\includegraphics[width=\columnwidth]{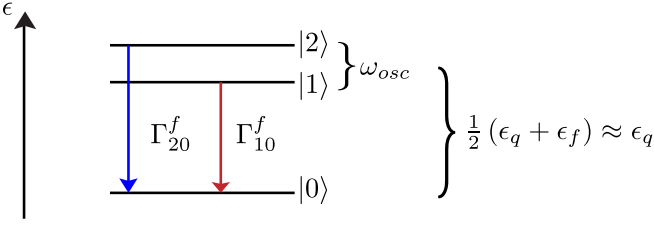}
		\caption{Level-structure of the coupled qubit-fluctuator system in the simplest case where only the TLF couples to a heat bath. 
			The rates $\Gamma_{10}^f$ and $\Gamma_{20}^f$ 	in Eq.~(\ref{eq:GammaT}) lead from levels $\ket{1}$ and $\ket{2}$ 
			respectively to the ground state $\ket{0}$. 
			The excited state $\ket{3}$ is not included in this illustration.}
		\label{fig:LevelsRates}
	\end{figure}

	For the purely decaying part we find
	\begin{eqnarray}
		a_{\downarrow, 1} = 2 \sin^4\frac{\xi}{2}\,, &a_{\downarrow, 2} = 2 \cos^4\frac{\xi}{2}\,, \nonumber \\
			\Gamma_{\downarrow, 1} = \Gamma_{10}^f = \cos^2\frac{\xi}{2} \: \gamma_1^f\,, &
			\Gamma_{\downarrow, 2} = \Gamma_{20}^f = \sin^2\frac{\xi}{2} \: \gamma_1^f \,,
		\label{eq:GammaT}
	\end{eqnarray} 
	where $\Gamma_{10}^f$ and $\Gamma_{20}^f$ are the rates with which the states $\ket{1}$ and $\ket{2}$ decay into the ground state 
	$\ket{0}$ (cf.~Fig.~\ref{fig:LevelsRates}).
	
	As we can see, the decay law for $\mean{\sigma_z(t)} - \mean{\sigma_z}_\infty$ is given by a sum of 
	several exponents. It is sometimes useful, e.g., for comparison with experiments where no fitting to 
	a specific decay law was performed, to define a single decay rate for the whole process.
	If a function $f(t)$ decays from $f(t=0)=a$ to $f(t\rightarrow\infty)=0$, we can define the single decay
	rate $\Gamma$ from $\int_0^\infty f(t)dt = a/\Gamma$. 

	We can introduce the single decay rate in two different ways, either effectively averaging over the oscillations 
	or including all parts and describing the envelope curve (cf. Fig. \ref{fig:SigmaMean}). 
	In the first case, averaging over the oscillations, we choose 
	$f(t)=a_{\downarrow, 1} \: e^{-\Gamma_{\downarrow, 1}\: t} + a_{\downarrow, 2} \: e^{-\Gamma_{\downarrow, 2}\: t}$
	and obtain for the amplitude and the decay rate
	\begin{eqnarray}
		a_{av} &=& a_{\downarrow, 1} + a_{\downarrow, 2} = 1 + \frac{\delta\omega^2}{v_\perp^2 + \delta\omega^2}
		\label{eq:AavA}\ ,\\
		\Gamma_{av} &=& \frac{a_{\downarrow, 1} + a_{\downarrow, 2}}
			{\frac{a_{\downarrow, 1}}{\Gamma_{\downarrow, 1}} + \frac{a_{\downarrow, 2}}{\Gamma_{\downarrow, 2}}} \nonumber\\
		&=& \frac{1}{2}\: \gamma_1^f\: \frac{v_\perp^2(v_\perp^2+2\delta\omega^2)}{v_\perp^4+5v_\perp^2\delta\omega^2+
			4\delta\omega^4}\ .
		\label{eq:GavA}
	\end{eqnarray}
	This gives a quasi-Lorentzian line shape of $\Gamma_{av}(\delta\omega)$ with the width of the order of the 
	coupling $v_\perp$ and the maximum value at resonance of
	$\Gamma_{av}(\delta\omega=0)= \frac{1}{2}\: \gamma_1^f$. 

	\begin{figure}[ht]
		\includegraphics[width=\columnwidth]{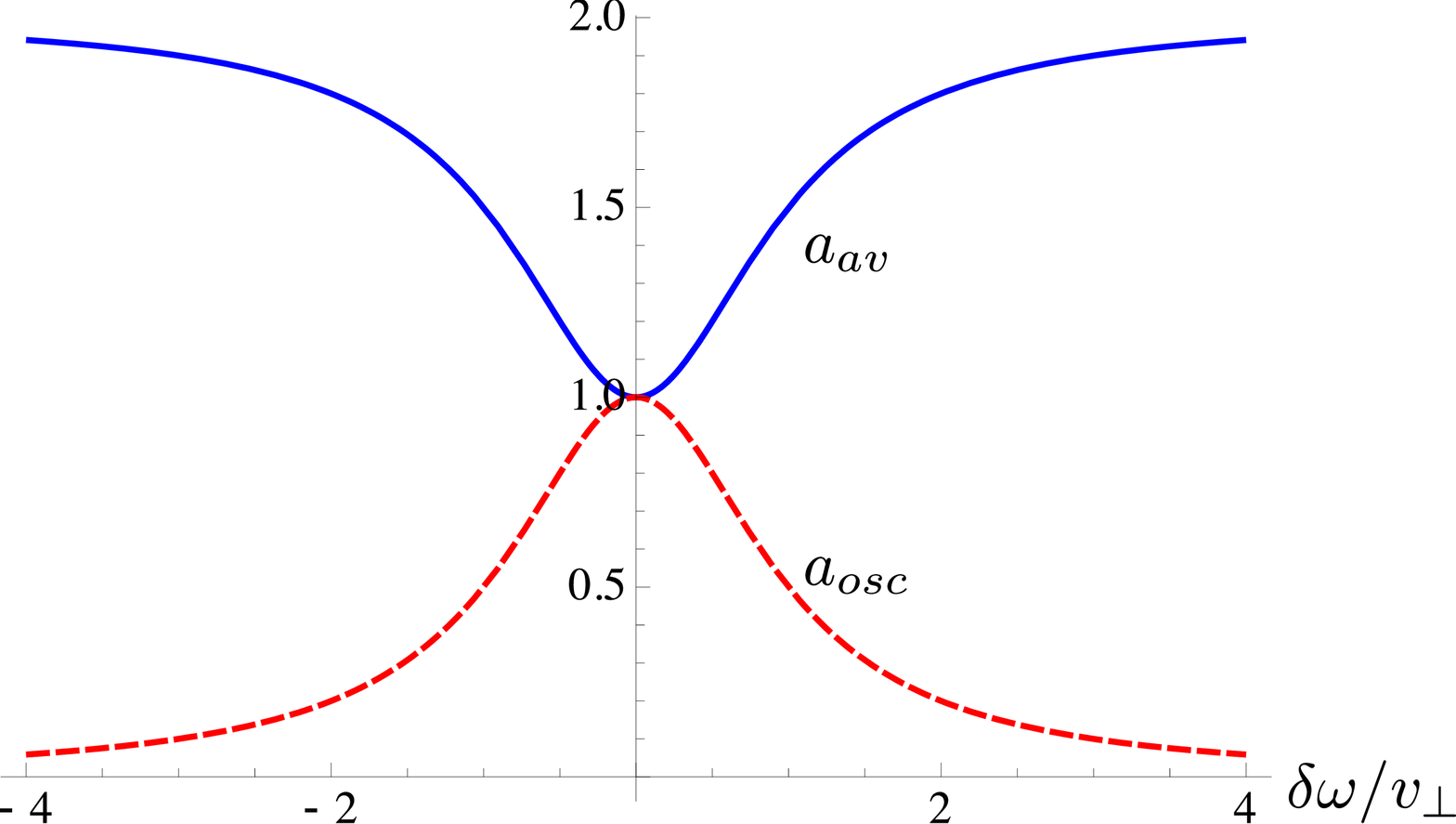}
		\includegraphics[width=\columnwidth]{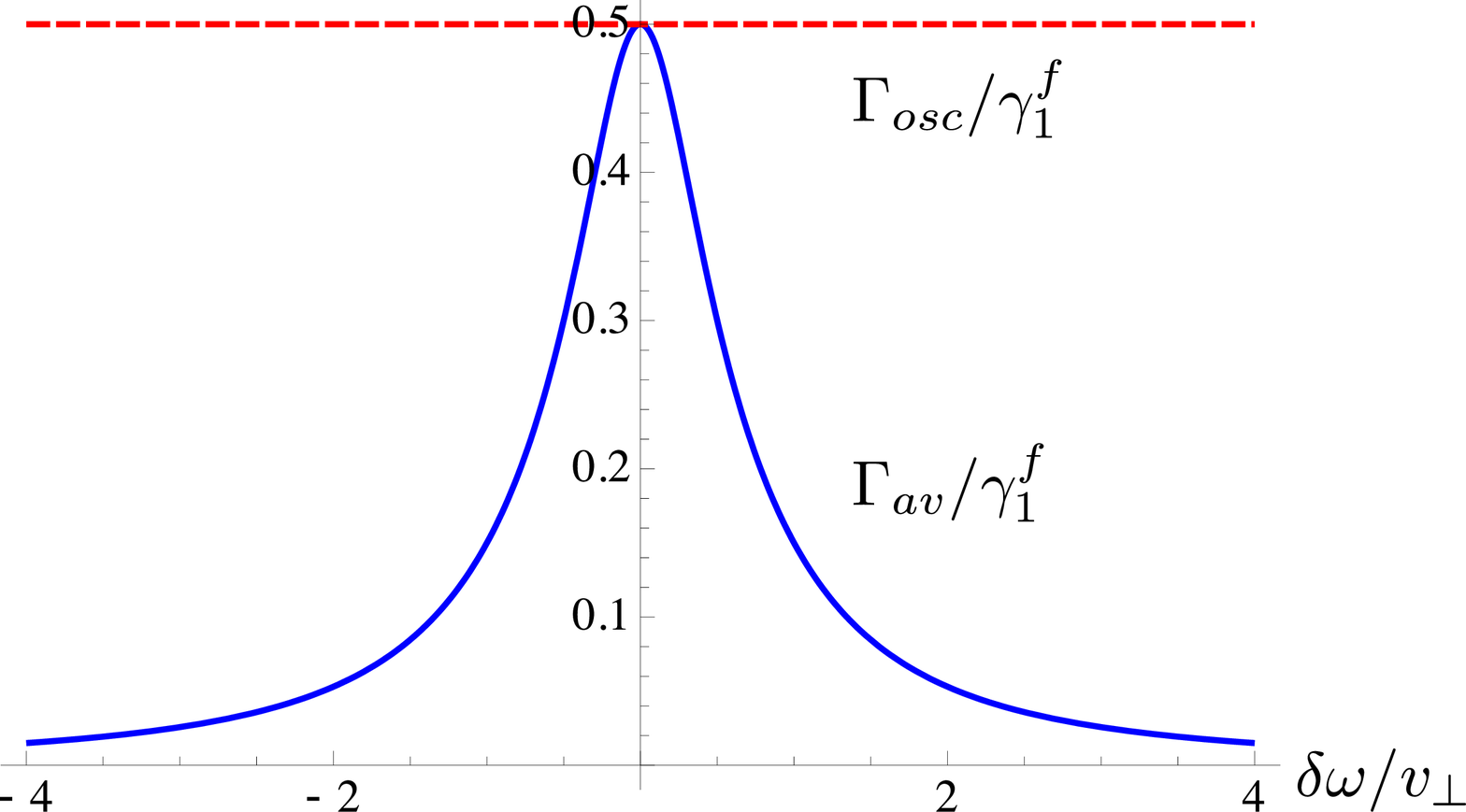}
		\caption{Amplitudes $a$ and rates $\Gamma$ for the decay of the oscillating (dashed red) and purely decaying (solid blue) part 
			of the qubit's $\mean{\sigma_{z}}$ as a function of the detuning $\delta\omega$ between the qubit and fluctuator. 
			The detuning is taken in units of the coupling $v_{\perp}$.}
		\label{fig:RatesAv}
	\end{figure}
	Figure~\ref{fig:RatesAv} shows the amplitudes and rates characterizing the decay of the oscillations (dashed red) and of the purely decaying part 
	(solid blue) of the qubits $\mean{\sigma_{z}}$. 

	To describe the envelope we choose $f(t)=a_{\downarrow, 1}
			\: e^{-\Gamma_{\downarrow, 1}\: t} + a_{\downarrow, 2} \: e^{-\Gamma_{\downarrow, 2}\: t} +
			a_{osc}\: e^{-\Gamma_{osc}\: t}$ (Fig. \ref{fig:SigmaMean}). This gives 
	\begin{eqnarray}
		\Gamma_{env} &=& \frac{a_{\downarrow, 1} + a_{\downarrow, 2} + a_{osc}}
			{\frac{a_{\downarrow, 1}}{\Gamma_{\downarrow, 1}} + \frac{a_{\downarrow, 2}}{\Gamma_{\downarrow, 2}} + 
			\frac{a_{osc}}{\Gamma{osc}}} \nonumber\\
		&=& \frac{1}{2}\: \gamma_1^f\: \frac{2v_\perp^2\: (v_\perp^2 + \delta\omega^2)}{2v_\perp^4 + 5v_\perp^2\: \delta
			\omega^2 + 4 \delta\omega^4}\ ,
		\label{eq:GenvA}			
	\end{eqnarray}
	with amplitude $a_{env} = a_{\downarrow, 1} + a_{\downarrow, 2} + a_{osc} = 2$. 
	This again gives a quasi-Lorentzian peak with height $\Gamma_{env}(\delta\omega=0)= \frac{1}{2}\: \gamma_1^f$ 
	and width similar to that of Eq.~(\ref{eq:GavA}).
	
	The results presented above are valid in the regime 
	when the coupling between the qubit and the TLF is stronger then the decay rates due to the interaction
	with the bath, $v_\perp \gg \gamma_1^f$. In the opposite limit the Golden rule results hold, with the 
	relaxation rate $\sim v_\perp^2/\gamma_1^f$ (cf. Refs.~\onlinecite{Paladino02,Makhlin03,Grishin05}).

\subsection{\label{sub:1TLFb}General coupling}

	We now provide the results for the general case, when both the qubit and TLF are coupled to heat baths.	
	The coupling in Eq.~(\ref{eq:H}) is given by
	\begin{eqnarray}
		\hat{H}_{\rm Bath} &=& \frac{1}{2} \left( \beta_{f,\parallel}\: \tau_z\cdot X_{f,\parallel} + \beta_{f,\perp}\: \tau_x \cdot X_{f,\perp} \right) \nonumber \\
			& + & \frac{1}{2} \left( \beta_{q,\parallel}\: \sigma_z\cdot X_{q,\parallel} + \beta_{q,\perp}\: \sigma_x \cdot X_{q, \perp}\right)\ .
		\label{eq:Hi2} 
	\end{eqnarray}
	It includes both transverse ($\perp$) and longitudinal ($\parallel$) coupling for both the qubit and the fluctuator.
	The temperature is still assumed to be well below the level splitting $\epsilon_q \approx \epsilon_f$ so that we can neglect excitation
	processes from the ground state.

	We specify now the main ingredients of the Bloch-Redfield tensor of the problems.
	As in Eq.~(\ref{eq:GammaT}) the relaxation rates from the states $\ket{1}$ and $\ket{2}$
	to the ground state due to the transverse coupling of the fluctuator are given by
	\begin{eqnarray}
		&\Gamma_{10}^f = \cos^2\frac{\xi}{2}\: \gamma_1^f\,,
		\qquad
		\Gamma_{20}^f = \sin^2\frac{\xi}{2}\: \gamma_1^f \,,&\nonumber\\
		&\gamma_1^f = \frac{1}{4}\: \beta_{f,\perp}^2\: C_{f,\perp}(\omega\approx\epsilon_q)\,.&
	\end{eqnarray}
	Similarly the transverse qubit coupling 
	gives rise to new rates,
	\begin{eqnarray}
		&\Gamma_{10}^q = \sin^2\frac{\xi}{2}\: \gamma_1^q\,,
		\qquad
		\Gamma_{20}^q = \cos^2\frac{\xi}{2}\: \gamma_1^q\,,& \nonumber\\
		&\gamma_1^q = \frac{1}{4}\: \beta_{q,\perp}^2\: C_{q,\perp}(\omega\approx\epsilon_q)\,.&
		\label{eq:gammaQ}
	\end{eqnarray}

	The longitudinal coupling to the baths, $\propto \sigma_z, \tau_z$, gives two types of additional 
	rates in the Redfield tensor, a pure dephasing rate, $\Gamma_\varphi$,
	and the transition rates between the states $\ket{1}$ and $\ket{2}$,
	\begin{eqnarray*}
		\Gamma_\varphi^f &=& \cos^2\xi\: \gamma_\varphi^f\\
		\Gamma_{12}^f &=& \frac{1}{4}\: \beta_{f,\parallel}^2\: \sin^2{\xi}\: C_{f,\parallel}\left( -\omega_{osc} \right)\\
		\Gamma_{21}^f &=& \frac{1}{4}\: \beta_{f,\parallel}^2\: \sin^2{\xi}\: C_{f,\parallel}\left( \omega_{osc} \right)\ ,
	\end{eqnarray*}
	where $\gamma_\varphi^f$ is the pure dephasing rate of the TLF:
	\begin{equation}
		\gamma_\varphi^f = \frac{1}{2}\: \beta_{f,\parallel}^2\: S_{f,\parallel}\left( \omega=0 \right)\ .
		\label{eq:gammaTD}
	\end{equation}
	Here $S_{f,\parallel}(\omega)=\frac{1}{2}\left[ C_{f,\parallel}(\omega) + C_{f,\parallel}(-\omega) \right]$ is the symmetrized 
	correlator.

\begin{figure}[ht]
		\includegraphics[width=\columnwidth]{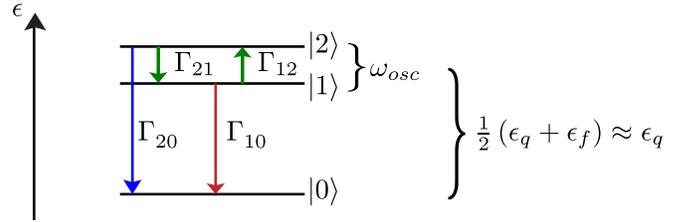}
		\caption{Illustration of the relevant transition processes in the general case of arbitrary coupling of the qubit and fluctuator to a heat bath.
			In addition to the transitions from the central levels $\ket{1}$, $\ket{2}$ to the ground state $\ket{0}$ with the rates $\Gamma_{10}$ 
			and $\Gamma_{20}$, we now also have transitions between the two central levels with the rates $\Gamma_{12}$ and $\Gamma_{21}$.
			The excited state $\ket{3}$ is again omitted in the illustration.}
		\label{fig:LevelsGeneral}
	\end{figure}

	Similarly, the rates due to the qubit's longitudinal coupling to the bath are given by
	\begin{eqnarray}
		\Gamma_\varphi^q &=& \cos^2{\xi}\: \gamma_\varphi^q, \nonumber\\
		\Gamma_{12}^q &=& \frac{1}{4}\: \beta_{q,\parallel}^2\: \sin^2{\xi}\: C_{q,\parallel}\left( -\omega_{osc} \right), \nonumber\\
		\Gamma_{21}^q &=& \frac{1}{4}\: \beta_{q,\parallel}^2\: \sin^2{\xi}\: C_{q,\parallel}\left( \omega_{osc} \right), \nonumber\\
		\gamma_\varphi^q &=& \frac{1}{2}\: \beta_{q,\parallel}^2\: S_{q,\parallel}\left( \omega=0 \right).
		\label{eq:gammaQD}
	\end{eqnarray}
	Figure~\ref{fig:LevelsGeneral} gives an illustration of the processes involved in the formation of the Redfield tensor.
	
	For $\mean{\sigma_z(t)}$ we again obtain the decay law, Eq. (\ref{eq:SigmaA}). The amplitude 
	of the oscillating part, $a_{osc}$, is still given by Eq.~(\ref{eq:aOscA}). The decay rate 
	of the oscillations is, however, modified:			
	\begin{equation}
		\Gamma_{osc} = \frac{1}{2} \left( \Gamma_1 + \Gamma_{12} + \Gamma_{21}\right) + \Gamma_\varphi.
		\label{eq:gammaOscB}
	\end{equation}
	The rates without a superscript represent the sum of the respective rates for the qubit and TLF:
	\begin{eqnarray*}
		\Gamma_1 = \gamma_1^f + \gamma_1^q, & \quad & \Gamma_\varphi = \Gamma_\varphi^f + \Gamma_\varphi^q, \nonumber\\
		\Gamma_{12} = \Gamma_{12}^f + \Gamma_{12}^q, & \quad & \Gamma_{21} = \Gamma_{21}^f + \Gamma_{21}^q. \nonumber
	\end{eqnarray*}

	The purely decaying part is given by a slightly more complicated expression. Defining
	\begin{eqnarray*}
		A &=& \Gamma_{10} + \Gamma_{12} \:\:\:\:\ , \:\:\:\:\: B = \Gamma_{20} + \Gamma_{21}\ ,\\
		C &=& \sqrt{\left( A-B \right)^2 + 4\Gamma_{12}\Gamma_{21}} \ ,
	\end{eqnarray*}	
	we obtain
	\begin{eqnarray*}
		a_{\downarrow, 1/2} &=& \frac{1}{2}\: \left( 1+\cos^2{\xi} \right) \nonumber\\
			& \mp & \frac{2\: \cos{\xi}\: (A - B) + \left( \Gamma_{12} + \Gamma_{21} \right) \sin^2{\xi}}{2\: C}\\
		\Gamma_{\downarrow, 1/2} &=& \frac{1}{2}\: \left( A + B \pm C \right)
	\end{eqnarray*}			
	In the limit $\beta_{q,\parallel}=\beta_{f,\parallel}=\beta_{q,\perp}=0$, we reproduce the results of the previous section. 

	The decay of the average is again characterized by
	\begin{eqnarray}
		a_{av} &=& a_{\downarrow, 1} + a_{\downarrow, 2} = 1 + \frac{\delta\omega^2}{v_\perp^2 + \delta\omega^2}
		\label{eq:AavAb}\ ,\\
		\Gamma_{av} &=& \frac{a_{\downarrow, 1} + a_{\downarrow, 2}}
			{\frac{a_{\downarrow, 1}}{\Gamma_{\downarrow, 1}} + \frac{a_{\downarrow, 2}}{\Gamma_{\downarrow, 2}}} \ .
		\label{eq:GavAb}
	\end{eqnarray}
	We work in the experimentally relevant limit $\omega_{osc}\ll T$. Then we obtain
	\begin{eqnarray*}
		\Gamma_{12}^f &=& \Gamma_{21}^f = \sin^2{\xi}\: \Gamma_v^f\ ,\\
		\Gamma_{12}^q &=& \Gamma_{21}^q = \sin^2{\xi}\: \Gamma_v^q\ ,
	\end{eqnarray*}
	where
	\begin{eqnarray}
		\Gamma_v^f &\equiv& \frac{1}{4}\: \beta_{f,\parallel}^2\: S_{f,\parallel}\left(\omega_{osc} \right)\ , \nonumber\\
		\Gamma_v^q &\equiv& \frac{1}{4}\: \beta_{q,\parallel}^2\: S_{q,\parallel}\left(\omega_{osc} \right) \ .
	\end{eqnarray}
	The decay rate of the average then reads
	\begin{equation}
		\Gamma_{av} = \frac{v_\perp^2 + 2\delta\omega^2}{2 \left( v_\perp^2 + \delta\omega^2 \right)}\: 
			\left\{\Gamma_1 - \frac{4 {\gamma_1^f}^2\: \delta\omega^2}{v_\perp^2 \left( \Gamma_1+4\Gamma_v \right) + 
			4 \gamma_1^f\: \delta\omega^2} \right\}\ ,
		\label{eq:GavB2}
	\end{equation}
	where $\Gamma_v = \Gamma_v^f + \Gamma_v^q$. 

\begin{figure}[ht]
		\includegraphics[width=\columnwidth]{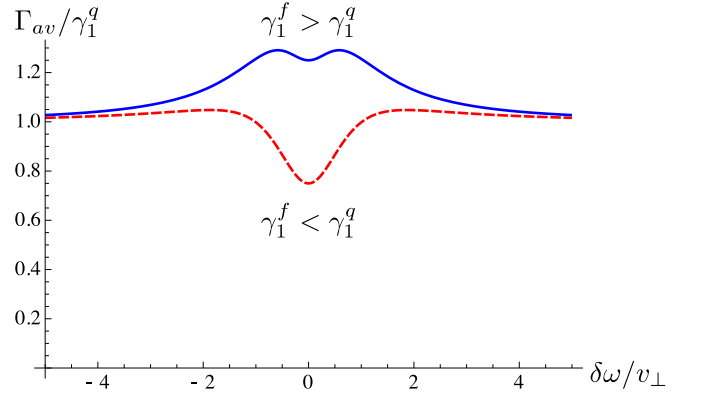}
		\caption{$\Gamma_{av}$ as a function of the detuning $\delta\omega$ 
			(rates in units of the qubit's relaxation rate $\gamma_{1}^q$, detuning in units of the coupling strength $v_{\perp}$) 
			for the general case when the TLFs relaxation rate $\gamma_1^f$ is 
			higher (solid blue, $\gamma_{1}^f=1.5 \gamma_{1}^q$) / lower (dashed red, $\gamma_{1}^f=0.5 \gamma_{1}^q$) 
			than the qubit's relaxation rate $\gamma_1^q$.
			The parameters in this plot are $\Gamma_v^q=\Gamma_v^f=0.3$, $v_{\perp}=5$ (in units of $\gamma_{1}^q$).}
		\label{fig:GammaAv2}
	\end{figure}

	At resonance the resulting relaxation rate $\Gamma_{av}$ is the mean of the decay rates of the qubit and TLF. 
	Thus, if the TLF relaxes slower than the qubit (as was the case in Ref.~\onlinecite{Neeley08}), $\Gamma_{av}$ decreases.
	Figure~\ref{fig:GammaAv2} shows the average decay rate $\Gamma_{av}$ for the two cases with $\gamma_{1}^f$ 
	bigger (solid blue) and smaller (dotted red) than $\gamma_{1}^q$. 
	The double-peaked structure in the first case is due to the contribution $\Gamma_v$ of
	the longitudinal coupling to the baths. Exactly in resonance and far away from resonance 
	the effect of $\Gamma_v$ vanishes, 
	while for $\delta\omega\sim v_{\perp}$ it produces somewhat faster relaxation.
	
\subsubsection*{\label{ssub:NonMarkov}Non-Markovian effects}

	In the previous section we have shown [see Eq.~(\ref{eq:gammaOscB})] that pure dephasing affects 
	only the decay rate of the oscillations of $\mean{\sigma_z}$ .
	The result, Eq. (\ref{eq:gammaOscB}), is valid for a short-correlated (Markovian) environment. 	
	More specifically the relations $\Gamma_\varphi^q = \cos^2{\xi}\: \gamma_\varphi^q$ with $\gamma_\varphi^q = \frac{1}{2}\: 
	\beta_{q,\parallel}^2\: S_{q,\parallel}\left( \omega=0 \right)$ and $\Gamma_\varphi^f = \cos^2{\xi}\: \gamma_\varphi^f$ 
	with $\gamma_\varphi^f = \frac{1}{2}\: \beta_{f,\parallel}^2\: S_{f,\parallel}\left( \omega=0 \right)$ are valid 
	only in the Markovian case. The generalization of these results to the case of non-Markovian noise, e.g., $1/f$ noise, 
	is straightforward~\cite{Ithier05}. Assuming that at low frequencies $S_{q,\parallel}=A_{q,\parallel}/\omega$ and 
	$S_{f,\parallel}=A_{f,\parallel}/\omega$, we obtain 
	\begin{equation}
		\begin{split}
			\mean{\sigma_z(t)} =& \mean{\sigma_z}_\infty + a_{\downarrow, 1}
			\: e^{-\Gamma_{\downarrow, 1}\: t} + a_{\downarrow, 2} \: e^{-\Gamma_{\downarrow, 2}\: t}\\
			+& a_{osc}\: \cos{\left( \omega_{osc}t \right)}\: f_{1/f}(t)\: e^{-\Gamma_{osc}'\: t}\ ,
		\end{split}
		\label{eq:SigmaA1overf}
	\end{equation}
	where $\ln f_{1/f}(t) \sim - t^2\: \cos^2{\xi}\:\left( \beta^2_{q,\parallel} A_{q,\parallel} +\beta^2_{f,\parallel} A_{f,\parallel} \right)$ and 
	$\Gamma_{osc}'$ now includes only Markovian contributions.

	In resonance, $\delta\omega=0$, we have $\cos{\xi}=0$ and therefore it seems at the first sight that the $1/f$ noise
	does not cause any dephasing. Yet, as shown in Ref.~\onlinecite{Makhlin04}, in this case the quadratic coupling becomes 
	relevant. The instantaneous splitting between the middle levels of the coupled qubit-TLF system,
	$\ket{1}$ and $\ket{2}$, is given by 
	\begin{eqnarray}
		\omega_{osc}(X_{q,\parallel},X_{f,\parallel}) &=& \sqrt{v_\perp^2 + (\beta_{q,\parallel} X_{q,\parallel} + 
			\beta_{f,\parallel} X_{f,\parallel})^2}\nonumber\\ 
			&\approx& v_\perp + \frac{1}{2} \frac{(\beta_{q,\parallel}
			X_{q,\parallel}+\beta_{f,\parallel} X_{f,\parallel})^2}{v_\perp}\ .\nonumber\\
	\end{eqnarray}
	This dependence produces a random phase between the states $\ket{1}$ and $\ket{2}$ and, as a result, 
	additional decay of the oscillations of $\mean{\sigma_z}$. We refer the reader to Ref.~\onlinecite{Makhlin04} for 
	an analysis of the decay laws and times. 
	Thus slow ($1/f$) fluctuations make the decay of the coherent oscillations of $\mean{\sigma_z}$ faster without 
	considerably affecting the average relaxation rate $\Gamma_{av}$.
	
	For strong $1/f$ noise, thus, a situation arises in which the oscillations decay much faster than
	the rest of $\mean{\sigma_z}$. In experiments with insufficient resolution this may appear as a 
	fast initial decay from $\mean{\sigma_z}=1$ to $\mean{\sigma_z}=1-a_{osc}$ followed by 
	a slower decay with the rate $\Gamma_{av}$. 

\subsection{\label{sub:2TLF}Two two-level fluctuators}

	As a first step towards the analysis of the effect of many TLFs,
	we examine now the case when two fluctuators are simultaneously at resonance with the qubit.
	This situation in the weak coupling regime was considered, e.g., in Ref~\cite{Ashhab06}.
	In this regime the fluctuators act as independent channels of decoherence
	and thus the contributions from different TLF are additive.
	However, in the regime of strong coupling between the qubit and the fluctuators, which is the focus of this
	paper, we do not expect the decoherence effects of the two fluctuators to simply add up. 
	This means that the resulting 
	relaxation rate is not given by the sum of two single-fluctuator rates.
	The Hamiltonian of the problem reads
	\begin{equation}
		\hat{H} = -\frac{1}{2}\: \epsilon_q\: \sigma_z - 
			\frac{1}{2}\:\sum_{n=1}^2\: \epsilon_{f,n}\: \tau_{z,n} + \frac{1}{2}\: \sigma_x\: \sum_{n=1}^2 v_{\perp,n}\: \tau_{x,n} 
			+ \hat{H}_{\rm Bath} \ ,
		\label{eq:H2}
	\end{equation}
	where $\hat{H}_{\rm Bath}$ contains now the coupling of each of the fluctuators to its respective bath.
	In the regime of our interest, $\epsilon_q \approx \epsilon_{f,1}\approx\epsilon_{f,2} \gg v_{\perp,1},v_{\perp,2}$, 
	the spectrum splits into four parts. The ground state is well 
	approximated by $\ket{g\uparrow\uparrow}$.
	Analogously, the highest excited state is close to $\ket{e\downarrow\downarrow}$. 
	The coupling $v_{\perp,n}$ is mainly relevant within two almost degenerate triplets. The first 
	triplet is spanned by the states with one excitation: 
	$\left\{ \ket{e\uparrow\uparrow}, \ket{g\downarrow\uparrow} , \ket{g\uparrow\downarrow}\right\}$. 
	In the second triplet, spanned by 
	$\left\{ \ket{g\downarrow\downarrow}, \ket{e\uparrow\downarrow}, \ket{e\downarrow\uparrow} \right\}$, 
	there are two excitations. At low temperatures and 
	for the initial state in which the qubit is excited and the fluctuators are in their ground states, only the first
	triplet and the global ground state are relevant. Within the first triplet the Hamiltonian reads
	\begin{equation*}
		\frac{1}{2}\: \left( \begin{array}{ccc}
			2 \epsilon_q & v_{\perp,1} & v_{\perp,2}\\
			v_{\perp,1} & 2 \epsilon_{f,1} & 0\\
			v_{\perp,2} & 0 & 2 \epsilon_{f,2}
			\end{array} \right)\ ,
	\end{equation*}
	where the energy is counted from the ground state.

	First, we consider the two fluctuators exactly in resonance with each other, 
	$\epsilon_{f,1}=\epsilon_{f,2}=\epsilon_f$, and approximately at resonance 
	with the qubit: $\epsilon_q\approx\epsilon_f$.
	We perform a rotation in the two-state subspace spanned by the states, where one of the TLF is excited, by applying the 
	unitary transformation
	\begin{equation}
		U=\left( \begin{array}{ccc}
			1 & 0 & 0\\
			0 & \phantom{-}\cos{\alpha} & \sin{\alpha}\\
			0 & -\sin{\alpha} & \cos{\alpha} 
			\end{array} \right)\ .
		\label{eq:2TLFRot}
	\end{equation}
	Choosing the angle $\alpha=\arccos{\frac{v_{\perp,1}}{\sqrt{v_{\perp,1}^2+v_{\perp,2}^2}}}$, we arrive at 
	the transformed Hamiltonian
	\begin{equation*}
		\frac{1}{2}\: \left( \begin{array}{ccc}
			2 \epsilon_q & \sqrt{v_{\perp,1}^2+v_{\perp,2}^2} & 0\\
			\sqrt{v_{\perp,1}^2+v_{\perp,2}^2} & 2 \epsilon_{f} & 0\\
			0 & 0 & 2 \epsilon_{f}
			\end{array} \right)\ .
		\label{eq:H2Trafo}
	\end{equation*}
	
\begin{figure}[ht]
		\includegraphics[width=\columnwidth]{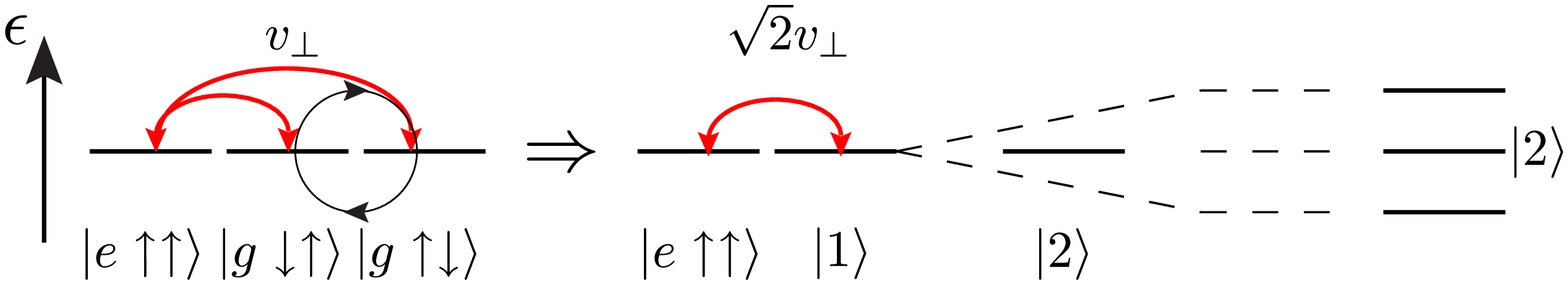}
		\caption{Illustration of the performed transformation in the one-excitation subspace of the Hamiltonian for the case of equal couplings
			$v_{\perp,1} = v_{\perp,2} = v_{\perp}$.
			Applying the rotation (\ref{eq:2TLFRot}) we arrive first in a situation where the state $\ket{2}$ (the dark state) is completely decoupled from
			the other two states. The renormalized coupling then splits the remaining two states, giving a situation analogous to the 
			coupling to one TLF.}
		\label{fig:3Levels}
	\end{figure}

	Figure~\ref{fig:3Levels} gives an illustration of what happens. After the rotation [Eq. (\ref{eq:2TLFRot})] the qubit is coupled to only one
	effective state $\ket{1}$, whereas it is completely decoupled from the ``dark'' state $\ket{2}$.
	For symmetric coupling $v_{\perp,1} = v_{\perp,2}$ the states $\ket{1}$ and $\ket{2}$ are just symmetric and antisymmetric superpositions of 
	$\ket{g\downarrow\uparrow}$ and $\ket{g\uparrow\downarrow}$.
	Thus, the rotation demonstrates that the situation is equivalent to only one effective TLF coupled to the qubit with the coupling strength 
	\begin{equation*}
		\tilde{v}_{\perp}=\sqrt{v_{\perp,1}^2+v_{\perp,2}^2} \ .
	\end{equation*}
	Analyzing the coupling of the effective TLF to the dissipative baths (of both fluctuators), we conclude 
	that the effective TLF is characterized by the relaxation rate 		
	\begin{equation*}
		\tilde{\gamma}^{f}_{1}=\frac{1}{v_{\perp,1}^2 + v_{\perp,2}^2} 
			\left( v_{\perp,1}^2\: \gamma_{1}^{f,1} + v_{\perp,2}^2\: \gamma_{1}^{f,2} \right)\ .
	\end{equation*}
	At this point we can apply all the results of Sec.~\ref{sub:1TLFa} with $v_{\perp}$ replaced by the renormalized 
	$\tilde{v}_{\perp}$ and the relaxation 
	rate $\gamma^{f}_{1}$ replaced by $\tilde{\gamma}^{f}_{1}$. In particular, using the formulas (\ref{eq:AavA}) 
	and (\ref{eq:GavA}) we introduce
	\begin{eqnarray}
		\Gamma_{av}^{(2)}(\delta \omega) = \Gamma_{av}(\delta \omega)|_{{v}_{\perp}\rightarrow \tilde{v}_{\perp},
			{\gamma}^{f}_{1}\rightarrow\tilde{\gamma}^{f}_{1}} \ .
		\label{eq:Gav2}
	\end{eqnarray}		
	Here the superscript $(2)$ stands for coupling to two fluctuators. The function $\Gamma_{av}^{(2)}(\delta \omega)$
	is peaked around $\delta\omega=0$. The height and the width of the peak depend on the relations between 
	the coupling strengths and the relaxation rates of the two fluctuators. In the limiting cases of `clear domination', 
	e.g., for $v_{\perp,1}\gg v_{\perp,2}$ and $\gamma_1^{f_1} \gg \gamma_1^{f_2}$, everything is determined 
	by a single fluctuator. In the opposite limit of identical fluctuators, i.e., 
	for $v_{\perp,1}=v_{\perp,2}$ and $\gamma_1^{f_1} = \gamma_1^{f_2}$, 
	the height of the peak [Eq. (\ref{eq:Gav2})] is given by $\gamma_1^{f_1}/2 = \gamma_1^{f_2}/2$, exactly 
	as in the case of a single fluctuator. The width of the peak is, however, $\sqrt{2}$ times larger since 
	$\tilde{v}_{\perp} = \sqrt{2}v_{\perp,1}=\sqrt{2}v_{\perp,2}$. Clearly, the 
	relaxation rate of the qubit is not given by a sum of two relaxation rates due to the two fluctuators.

	If the fluctuators are not exactly in resonance, this result still holds as long as their detuning $\delta\epsilon_{f} = \epsilon_{f,1}-\epsilon_{f,2}$ 
	is smaller than the renormalized coupling $\tilde{v}_{\perp}$. For much larger detuning, the rate is given by the sum of two single-TLF contributions. 
	
\subsection{\label{ManyFluctuators} Many degenerate fluctuators}
			
	For a higher number of fluctuators in resonance, i.e., $\epsilon_{f,n} = \epsilon_{f}$, 
	the argument presented above is still valid, and the resulting 
	decoherence of the qubit's state is the same as for a single TLF with a renormalized coupling strength of 
	\begin{equation*}
		\tilde{v}_\perp=\sqrt{\sum_{n} v_{\perp,n}^2} 
	\end{equation*}
	and an effective TLF relaxation rate 
	\begin{equation*}
		\tilde{\gamma}_{1}^{f} = \frac{1}{\tilde{v}_{\perp}^2} \sum_{n} v_{\perp,n}^2\: \gamma_{1}^{f,n}\ .
	\end{equation*}
	It should be stressed that this equivalence holds only within the one-excitation subspace of the system.
	 
	If the system with many fluctuators is excited more than once it is no longer 
	equivalent to a system with one effective TLF. Multiple excitation 
	could be achieved e.g., by following the procedure used in Ref.~\cite{Hofheinz2009} or that of 
	Ref.~\cite{Neeley08} repeatedly, i.e., exciting the qubit while out of resonance, transferring 
	its state to the TLFs, exciting the qubit again and so forth.
	The simplest case is when all the fluctuators have equal couplings to the qubit $v_{\perp,n} = v_{\perp}$.
	The system's Hamiltonian reads then
	\begin{equation}
		H = -\frac{1}{2}\,\epsilon_q \sigma_z - \epsilon_f S_z + v_{\perp} \sigma_x S_x\ ,
	\end{equation}
	where $S_\alpha \equiv (1/2)\sum_n \tau_{\alpha,n}$.

	For procedures of the type used in Ref.~\cite{Neeley08,Hofheinz2009}, i.e., when only the qubit can be 
	addressed, the TLFs will remain in the spin representation of $S_\alpha$ in which they were 
	originally prepared. If the TLFs are all initially in their ground states, the accessible part of 
	the Hilbert space is that of a qubit coupled to a spin $N/2$, where $N$ is the number of TLFs. 
	For the procedure similar to that of \cite{Hofheinz2009} the oscillation periods with $k$ excitations would be 
	given by $2\pi/\tilde{v}_{\perp,k}$, where $\tilde{v}_{\perp,k} = \sqrt{k \: \left( N+1-k \right)} \cdot v_{\perp}$.
					
\section{\label{sec:Statistics}Collection of fluctuators}

We now analyze decoherence of a qubit due to multiple TLFs. For this purpose, we introduce an ensemble of TLFs with energy splittings $\epsilon_{f,n}$, distributed randomly. For each fluctuator $n$ we assume a uniform distribution of its energy splitting $\epsilon_{f,n}$ in a wide interval $\Delta E$, with probability density $p_n = 1/\Delta E$. The overall density of fluctuators is given by $\nu_0\equiv N/\Delta E$, where $N$ is the total number of fluctuators in the interval $\Delta E$. For simplicity we assume all the fluctuators to have the 
same coupling to the qubit $v_\perp$ and the same relaxation rate $\gamma_1^f$. 
The interval is much wider than a single peak, $\Delta E \gg v_{\perp}$, and 
the total number of TLFs in the interval is $N=\nu_0\: \Delta E\gg 1$.
	
We find that the physics is controlled by the dimensionless parameter $\bar{\nu} \equiv \nu_0\: v_{\perp}$.
For $\bar{\nu}\ll 1$ the probability for two fluctuators to be in resonance with each other is low. 
Once the qubit is in resonance with one of the TLFs, the decay law of the qubit's $\langle \sigma_z \rangle$
takes the form (\ref{eq:SigmaA}). In this regime we take $\Gamma_{av}$ to characterize the decay.
We expect that in most situations the oscillations in Eq. (\ref{eq:SigmaA}) will decay fast due to the pure dephasing,
and one will observe a very fast partial (down to half an amplitude) decay of $\langle \sigma_z \rangle$ followed
by further decay with rate $\Gamma_{av}$. Thus the relaxation 
rate is given by a sum of many well separated peaks, each contributed by a single fluctuator. 
Since the positions of the peaks are random, we expect, for $\bar\nu\sim1$, a randomly looking dependence of the qubit's
relaxation rate on the qubit's energy splitting (and a collection of rare peaks for lower densities, $\bar\nu\ll1$).
To characterize the statistical properties, we determine in Sec.~\ref{sub:NoCorrections} the relaxation rate, averaged over realizations, and its variance. 

For larger $\bar{\nu} > 1$ the situation changes, as the peaks become dense, and the 
probability to have two or more fluctuators in resonance with each other is high. We conclude that 
it is not reasonable anymore to characterize the decay of $\langle \sigma_z \rangle$ by $\Gamma_{av}$. In this limit the coherent oscillations turn into much faster relaxation. The excitation energy is transferred from the qubit to the 
TLFs on a new, short time scale, $\sim (\bar{\nu} v_\perp)^{-1}$, which we now call the relaxation time. The energy remains in the TLFs for much longer time ($\sim 1/\gamma^f_1$) before it is released to the 
dissipative baths. Yet, if a strong enough fluctuation of the TLFs spectral density occurs, coherent oscillations
appear again. A set of TLFs almost at resonance with each other form an effective strongly coupled
fluctuator. The decay time of the oscillations is due to the background density of TLFs rather than due to the 
coupling to the baths. This could be an alternative explanations for the 
findings of Refs.~\onlinecite{Simmonds04,Cooper04,Neeley08}.
In Sec.~\ref{sub:HighDensities} we describe these two situations.

\subsection{\label{sub:NoCorrections}Independent fluctuators, $\bar\nu\ll1$. }
		
	In this regime the relaxation rate $\Gamma$ is given by a sum of single-fluctuator contributions.
	For a given realization of the ensemble we obtain
	\begin{equation}
		\Gamma(\epsilon_q) = \sum_n \Gamma_{av} (\epsilon_q-\epsilon_{f,n})\ .
		\label{eq:Ind_Fluct}
	\end{equation}
	Integrating over the TLF energy splittings $\epsilon_{f,n}$, we obtain the average relaxation rate
	\begin{equation}
		\mean{\Gamma}= \int\: d^N\!\epsilon\; p^{(N)} 
		\sum_n \Gamma_{av} (\epsilon_q-\epsilon_{f,n}) \sim \gamma_1^f\: \bar{\nu}
		\label{eq:Mean0}
	\end{equation}
	Here $d^N\epsilon=d\epsilon_{f,1}\dots\: d\epsilon_{f,N}$,
	and the probability distribution $p^{(N)} = \prod_n p_n$ is given by the product of 
	single-TLF distribution functions,
	$p_n=\frac{1}{\Delta E}=\frac{\nu_0}{N}$.

	We also find the variance 
	\begin{equation}
		\var{\Gamma^2} = \mean{\Gamma^2} - \mean{\Gamma}^2 \sim (\gamma_1^f)^2 \bar{\nu}\ .
		\label{eq:Var0}
	\end{equation} 
	Thus		
	\begin{equation}
		\frac{\var{\Gamma^2}}{\mean{\Gamma}^2} \sim \frac{1}{\bar{\nu}}
	\end{equation}
	This result can be expected. In the regime $\bar{\nu}\ll 1$ in each realization of the environment the function 
	$\Gamma(\epsilon_q)$ is a collection of rare peaks of height $\frac{1}{2}\gamma_1^f$ and width $v_\perp$. The 
	average value of $\Gamma$ is, thus, small, but the fluctuations are large. 
	As expected, the relative variance decreases as the effective density $\bar{\nu}$ increases.
		
	As we have seen in Sec.~\ref{sub:2TLF}, the contribution from two TLFs in resonance differs from the sum of two single-TLF contributions. 
	This effect leads to modifications of Eqs.~(\ref{eq:Mean0}) and (\ref{eq:Var0}) with the further increase in the spectral density $\bar\nu$ of the fluctuators. 
	The relaxation rate becomes lower than the result, Eq. (\ref{eq:Mean0}), 
	in the approximation of independent fluctuators, and the straightforward estimate gives:
	\begin{equation}
		\mean{\Gamma} \propto \gamma_1^f \:(\bar{\nu} - c_{1} \bar{\nu} ^2)\ ,
	\end{equation}
	Similarly, for the variance (which is close to the mean square) we find
	\begin{equation}
		\var{\Gamma^2} \propto (\gamma_1^f)^2 \:(\bar{\nu}- c_{2} \bar{\nu} ^2)\ .
	\end{equation} 
	Here both prefactors $c_1, c_2 \sim1$. For example, $c_1=2-\sqrt{2}, c_2\approx1.71$ in the rough approximation, 
	when we (i) account for correlations by using the rate [Eq. (\ref{eq:Gav2})] for two resonant TLFs to describe the joint effect of two fluctuators, 
	$n$ and $m$, in a certain range around resonance, i.e., when their energy splittings differ by less than the coupling strength, 
	$|\epsilon_{f,n}-\epsilon_{f,m}|< v_{\perp}$; and (ii) neglect correlations for larger detunings $|\epsilon_{f,n}-\epsilon_{f,m}|$.

	\begin{figure}[ht]
		\includegraphics[width=\columnwidth]{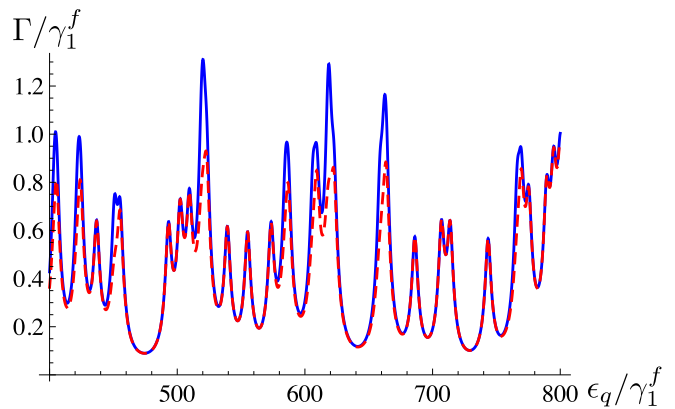}
		\caption{$\Gamma$ as a function of the qubit's level splitting $\epsilon_{q}$ in units of the TLF relaxation rate $\gamma_{1}^f$ 
			for one possible realization of the fluctuator distribution with $\bar{\nu} = 0.5$ and $v_{\perp}=5 \gamma_{1}^f$. 
			Solid blue line: approximation of independent fluctuators (\ref{eq:Ind_Fluct}). Dashed red line: with account for correlations, see text.
			The corrections are most pronounced in areas, where more than one fluctuator is in resonance 
			($\Gamma > \frac{1}{2} \gamma_{1}^f$).}
		\label{fig:GammaReal}
	\end{figure}

	Figure~\ref{fig:GammaReal} shows the relaxation rate $\Gamma$ for one possible realization of the TLF 
	distribution at $\bar{\nu} = 0.5$. The solid blue line corresponds to the approximation of independent 
	fluctuators [Eq. (\ref{eq:Ind_Fluct})], while the dashed red line is calculated using the approximation described. 
	
	We see, that the experimental data, where ``random'' behavior of the relaxation rate as a function of the 
	qubit's energy splitting was observed, could be consistent with the situation depicted in 
	Fig.~\ref{fig:GammaReal}, i.e, with $\bar\nu\sim 1$.

\subsection{\label{sub:HighDensities}Spectrally dense fluctuators, $\bar\nu\gg1$}
	
	For higher densities $\bar{\nu}$ the calculations above are no longer valid. In this section we discuss the limit of very high spectral densities, $\bar\nu\gg1$.
	In the following we restrict ourselves to the one-excitation subspace of the system and neglect the 
	couplings to the baths. Thus we consider the one-excitation subspace of the following Hamiltonian
	\begin{equation}
		\hat{H} = -\frac{1}{2}\: \epsilon_q\: \sigma_z - 
			\frac{1}{2}\:\sum_{n}\: \epsilon_{f,n}\: \tau_{z,n} + \frac{1}{2}\: \sigma_x\: \sum_{n} v_{\perp,n}\: \tau_{x,n}\ .
		\label{eq:Hpure}
	\end{equation}	
	\begin{figure}[ht]
		\includegraphics[width=\columnwidth]{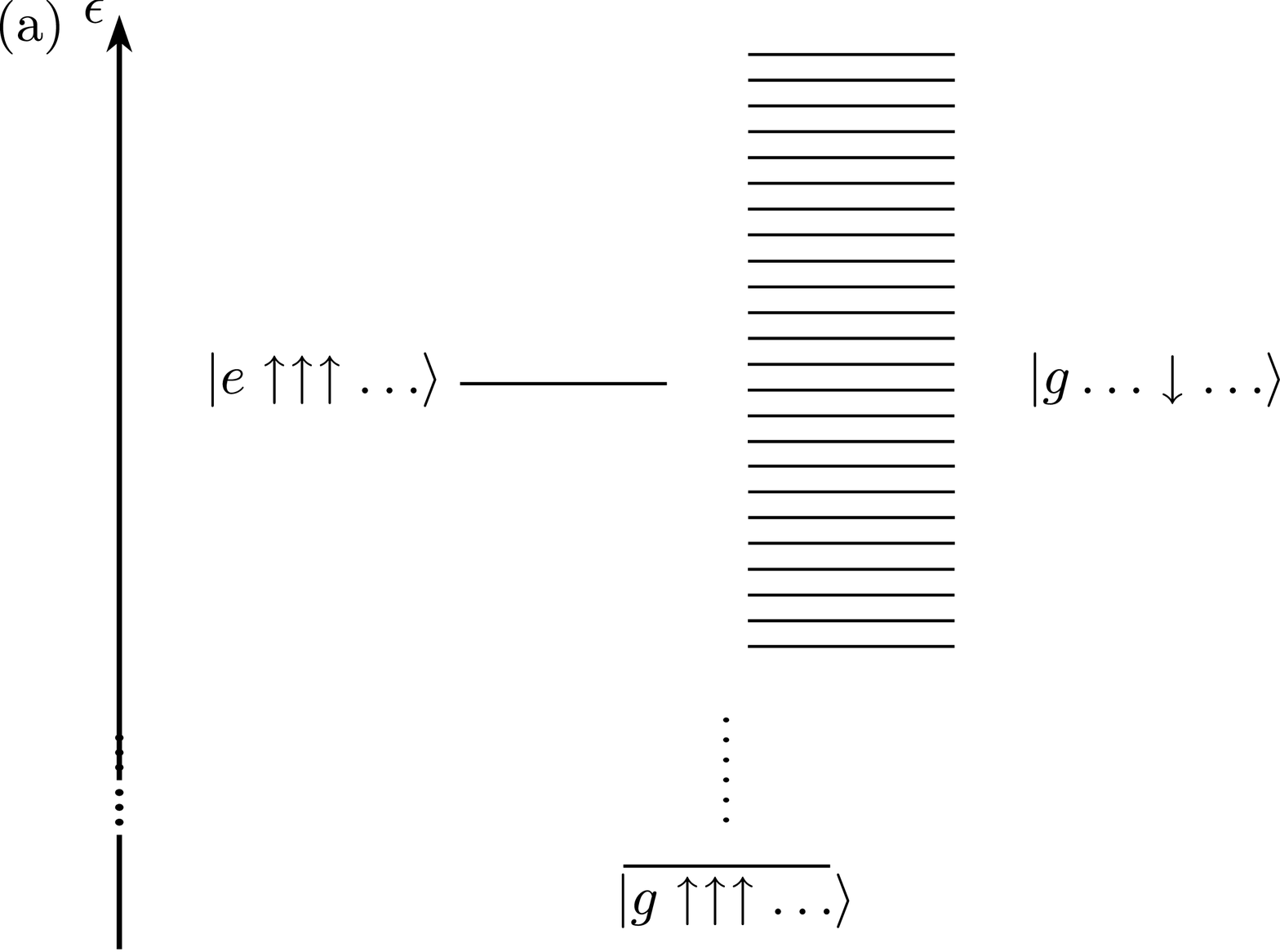}\vskip .5cm
		\includegraphics[width=\columnwidth]{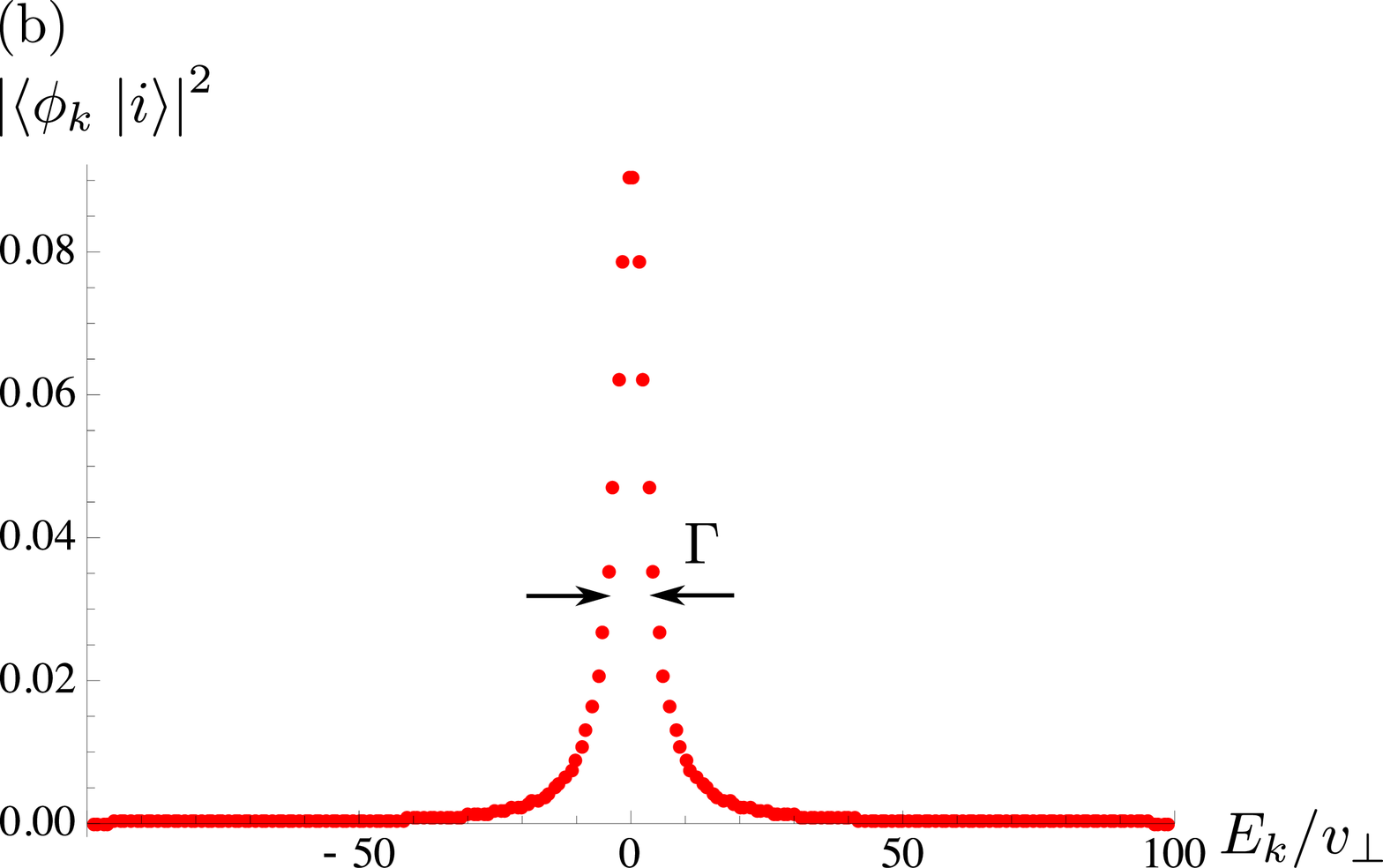}
		\caption{{\bf (a)} Level structure of a qubit coupled to a uniform spectral 
			distribution of fluctuators. The one-excitation subspace is shown. The state, where the qubit is excited, 
			$\ket{e\uparrow\uparrow\uparrow\ldots}$, is coupled with strength $v_{\perp}$ to all other levels in this subspace. 
			The ground state $\ket{g\uparrow\uparrow\uparrow\ldots}$ is energetically well separated 
			from the subspace with one excitation. {\bf (b)} Overlap of the initial state
			$\ket{i}=\ket{e\uparrow\uparrow\uparrow\ldots}$ with the eigenstates of the coupled system 
			$\ket{\phi_k}$ for a uniform distribution of TLFs with
			an effective density $\bar{\nu} = 1$ as depicted above. The energy is counted from the energy 
			of the initial state.}
		\label{fig:OverlapCont}
	\end{figure}
	Our purpose is to diagonalize the Hamiltonian in the one-excitation subspace and to find the overlap 
	of the initial state $\ket{i}$ (qubit excited, all fluctuators in the ground state) with the eigenstates 
	$\ket{\phi_k}$, labeled by an index $k$ and having the eigenenergies $E_k$. This allows us 
	to obtain the time evolution of the initial state:
	\begin{equation}
		\ket{ i (t) } = \sum_{k} \ket{\phi_{k}} \braket{\phi_{k}}{ i } e^{-i E_{k} t}\ .
		\label{eq:it}
	\end{equation}
		
	We begin with the case of a completely uniform spectral distribution.	
	This case is well known in quantum optics as the
	Wigner-Weisskopf theory~\cite{Weisskopf1930}. We obtain a Lorentzian shape of the overlap function.
	Figure~\ref{fig:OverlapCont} shows the overlap $\left| \braket{\phi_{k}}{i} \right|^2$ for an 
	effective density $\bar{\nu} = 1$ as a function of $E_k$. 
	We arrive at the probability amplitude to find the qubit still excited after time $t>0$:
	\begin{eqnarray}
		\braket{i}{i (t) } &=& \sum_k \left| \braket{\phi_{k}}{i} \right|^2 e^{-i E_k t}\nonumber\\
		&=&\int d E \sum_k \left| \braket{\phi_{k}}{i} \right|^2 \delta(E-E_k) e^{-i E t} 
		\nonumber\\
		&=& \int \frac{d E}{2\pi}\: \frac{\Gamma}{\left(\frac{\Gamma}{2}\right)^2 + E^2} e^{-i E t} = e^{-\frac{\Gamma}{2} t}\ ,
		\label{eq:iLorentz}
	\end{eqnarray}
	where $\Gamma =\frac{\pi}{2}\, \bar{\nu} \cdot v_{\perp}$.
	Thus the decay of the initially excited qubit in this situation is described by a simple 
	exponential decay, $|\braket{i}{i (t) }|^2 = e^{-\Gamma t}$.
	The width $\Gamma$ of the Lorentzian in Fig.~\ref{fig:OverlapCont} determines the decay rate of the excited state.

	Note, that we did not include here the coupling of either the qubit or the fluctuators to the 
	dissipative baths. These couplings will broaden each of the eigenstates by an amount 
	$\sim \gamma_1^f$. As long as this broadening is smaller than the resulting decay rate 
	$\Gamma =\frac{\pi}{2}\, \bar{\nu} \cdot v_{\perp}$ (for strongly coupled fluctuators ($v_\perp\gg\gamma_1^f$) and 
	$\bar{\nu} >1$ this is always the case), the dissipative broadening has little effect. Thus 
	the system of the coupled qubit and fluctuators remains for a long time ($\sim 1/\gamma_1^f$) 
	in the one-excitation subspace, but the qubit relaxes much faster and the energy resides in the 
	fluctuators.
	
	\begin{figure}[ht]
		\includegraphics[width=\columnwidth]{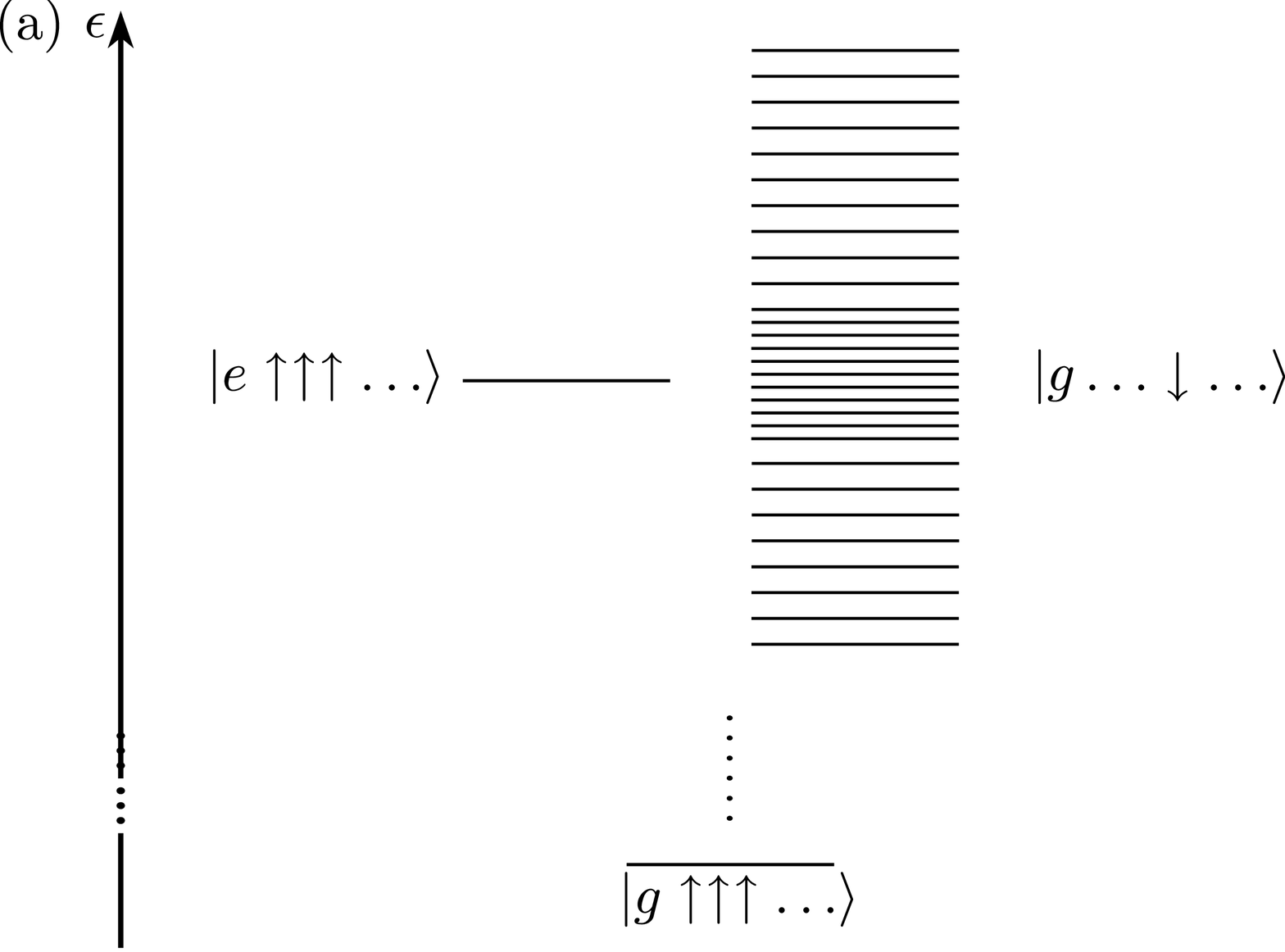}\vskip .5cm
		\includegraphics[width=\columnwidth]{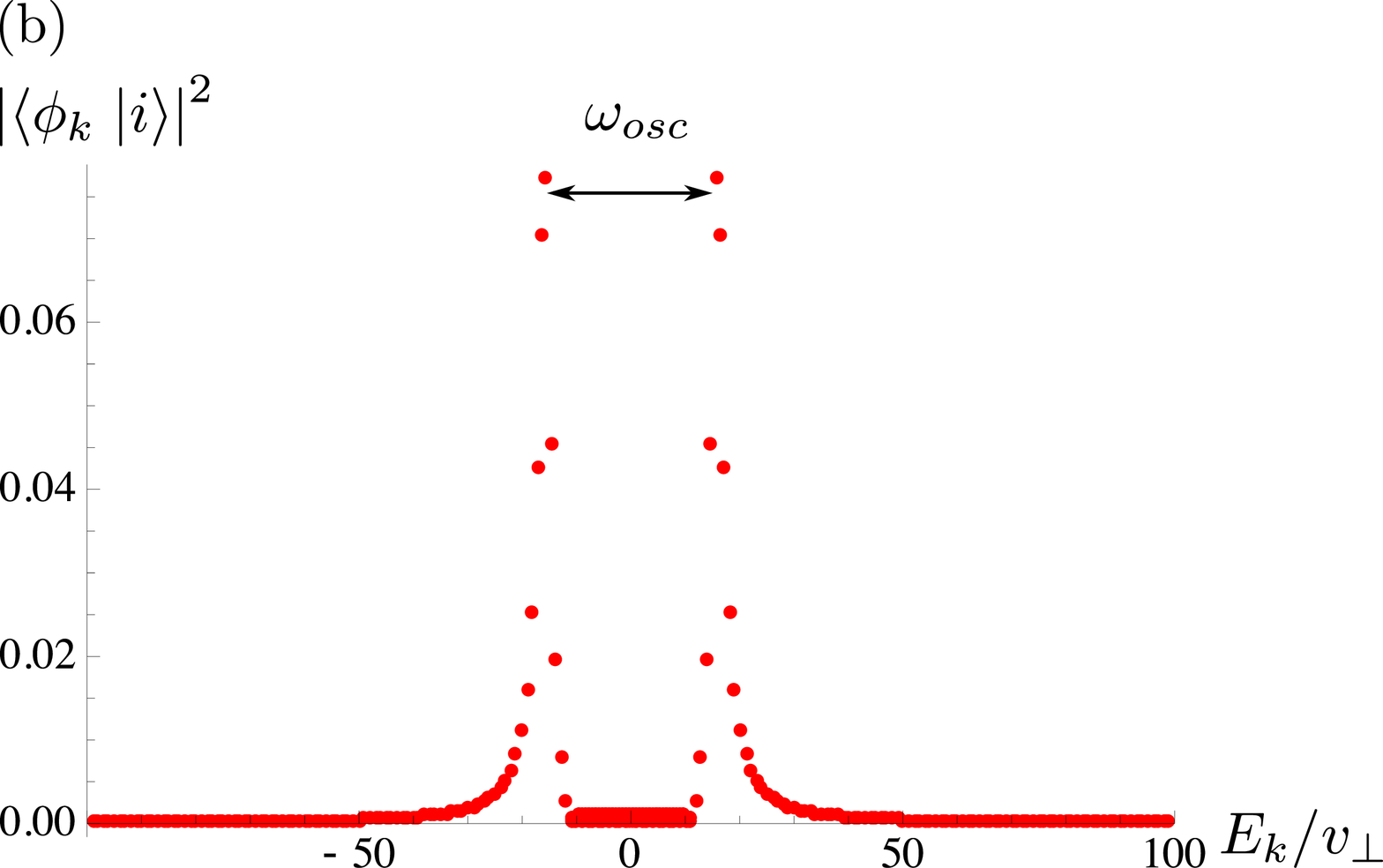}
		\caption{{\bf (a)} Level structure for a non-homogeneous spectral distribution of TLFs.
			At energies near the qubit's level splitting $\epsilon_{q}$ the density is increased. The state 
			$\ket{e\uparrow\uparrow\uparrow\ldots}$ is coupled equally to each level in the one-excitation subspace.
			{\bf (b)} Overlap of the initial state with the eigenstates of the coupled system 
			with a higher local density $\bar{\nu}_{local}=10$ between $\epsilon=-10$ and
			$\epsilon=10$ (in units of $v_\perp$). Out of resonance the effective density is $\bar{\nu}=1$.}
		\label{fig:OverlapLocal}
	\end{figure}

	Further, we analyze the situation with a large number of TLFs, whose energy splittings are accumulated near some value. 
	For instance, this behavior may originate from the microscopic nature of the fluctuators. As we have seen above, a collection of resonant TLFs is equivalent, 
	within the single-excitation subspace, to one effective TLF with a much stronger coupling to the qubit. 
	This results in two energy levels, separated by this new strong effective coupling constant. 
	As we discussed above, this may be the origin of the visible properties of strongly coupled TLFs. 
	To illustrate this setting, we show typical numerical results in Fig.~\ref{fig:OverlapLocal}. The data are shown for the situation,
	 where instead of many resonant TLFs we have a large collection of TLFs distributed in a certain energy range. 
	 On top of the homogeneous distribution with density $\bar{\nu}=1$, we
	assume, locally, a higher density ($\bar{\nu}=10$) of fluctuators with energies close to that of the qubit.
	We obtain a double-peak structure for the overlap $\left| \braket{\phi_{k}}{i} \right|^2$. 
	Performing again a calculation of Eq.~(\ref{eq:iLorentz}), we obtain oscillations with frequency 
	given by the energy splitting of the two peaks in Fig.~\ref{fig:OverlapLocal}. The widths of the peaks (set at this level by the local 
	density) determine the decay rate of the oscillations. If these peaks are wider than the dissipative broadening, the latter can be neglected.
	Thus, the effect of the fluctuator bath with strong density variations on the qubit is equivalent to that of
	a single TLF with an effective coupling strength $\tilde{v}_\perp$,
	much stronger than the couplings $v_{\perp,n}$ between the qubit and the individual physical TLFs.

\section{\label{sec:Conclusions} Conclusions}

In this paper we examined the effect of strongly coupled two-level fluctuators on the dissipative dynamics 
of a qubit. We have described the following phenomena:

(a) If the qubit and TLF are close to resonance, one should observe 
coherent oscillations between them~\cite{Neeley08}. We have analyzed the 
effects of dissipation and the relaxation rates. These results may be important in the studies of
single two-level fluctuator systems, e.g., with the idea to use them as quantum memory.

(b) The situation of a qubit coupled to several TLFs with degenerate level splittings is equivalent 
to the coupling of the qubit to a single effective TLF with a renormalized, stronger coupling strength.

(c) Collections of TLFs with spectral density $\bar{\nu}\sim 1$ could show a seemingly random 
dependence of the qubit's relaxation rate on its level splitting.

(d) In the dense limit, $\bar{\nu}\gg 1$, we conclude that uniform distributions 
of the TLF energies lead to the exponential relaxation of the qubit. 
Local strong fluctuations of the density in the energy distribution can, however, from the qubit's viewpoint appear as 
a single strongly coupled effective TLF.
We conclude that the signatures of strong coupling to single two-level systems often observed in qubit 
spectroscopy could well arise from weak coupling to many nearly resonant TLFs.
We emphasize that this equivalence holds only for the initial state, in which the 
qubit is excited and the fluctuators are not. 

\section*{Acknowledgments}

We acknowledge support from EuroSQIP, startup fund of the 
Rector of the Karlsruhe University (TH), INTAS, the Dynasty foundation, RFBR and the U.S. ARO under Contract No. W911NF-09-1-0336.
We want to thank J. Cole, J. Lisenfeld, M. Marthaler, G. Sch\"on, and A. Ustinov for valuable discussions.

\bibliography{TLS_paper} 

\end{document}